\documentclass[conference]{IEEEtran}
\IEEEoverridecommandlockouts
\usepackage{cite}
\usepackage{amsmath,amssymb,amsfonts}
\usepackage{algorithmic}
\usepackage{graphicx}
\usepackage{textcomp}
\newcommand{\color}[1]{\null}

\def\BibTeX{{\rm B\kern-.05em{\sc i\kern-.025em b}\kern-.08em
T\kern-.1667em\lower.7ex\hbox{E}\kern-.125emX}}
\usepackage{cite}

\DeclareGraphicsExtensions{.pdf,.jpg,.png,.jpeg}

\begin{document}
	
\title{~\vspace{-12mm}\\{\normalsize Author preprint February 2025;
to appear in IEEE PES General Meeting Austin TX USA July 2025 \copyright IEEE 2025}\\
\vspace{-1mm}
Resiliency metrics quantifying emergency response in a distribution system }

\author{\IEEEauthorblockN{Shikhar Pandey}
\IEEEauthorblockA{\textit{Commonwealth Edison Company} \\
Chicago, IL USA \\
shikhar.pandey@comed.com}
\and
\IEEEauthorblockN{Gowtham Kandaperumal*}
\IEEEauthorblockA{\textit{Commonwealth Edison Company} \\
Chicago, IL USA \\
Gowtham.Kandaperumal@comed.com}
\and
\IEEEauthorblockN{Arslan Ahmad}
\IEEEauthorblockA{\textit{Iowa State University} \\
Ames IA USA \\
arslan@iastate.edu}
\and
\IEEEauthorblockN{Ian Dobson}
\IEEEauthorblockA{\textit{Iowa State University} \\
Ames IA USA \\
dobson@iastate.edu}
\thanks{*Corresponding author.}
\thanks{The authors from Iowa State University gratefully acknowledge support from PSerc, Iowa State University EPRC, and NSF grants 2153163 and 2429602.}

}


\maketitle	

\begin{abstract}
The electric distribution system is a cornerstone of modern life, playing a critical role in the daily activities and well-being of individuals. As the world transitions toward a decarbonized future, where even mobility relies on electricity, ensuring the resilience of the grid becomes paramount. 
This paper introduces novel resilience metrics designed to equip utilities and stakeholders with actionable tools to assess performance during storm events. 
The metrics focus on emergency storm response and the resources required to improve customer service. The practical calculation of the metrics from historical utility data is demonstrated for multiple storm events. Additionally, the metrics’ improvement with added crews is estimated by “rerunning history” with faster restoration. By applying this resilience framework, utilities can enhance their restoration strategies and unlock potential cost savings, benefiting both providers and customers in an era of heightened energy dependency.  \end{abstract}


\section{Introduction}
\looseness=-1
Distribution system resilience can be divided into two key categories: system resilience and operational resilience. 
System resilience refers to the ability of the distribution grid to withstand extreme weather events while continuing to serve customers. 
Operational resilience, on the other hand, is a quantification of the restoration efforts, including resource deployment, grid automation performance, and the efficiency of the resources used to restore service to customers. Metrics that can assess both these categories are important to understand overall distribution system resilience. 

In existing literature, resilience metrics often focus heavily on the restoration of critical loads, such as hospitals and police stations. 
While restoring these essential services remains a priority, this paper introduces operational metrics aimed at optimizing the restoration of all loads, ensuring an effective and efficient recovery process across the entire grid.

When the distribution grid encounters extreme weather and experiences outages, the restoration process is immediately activated. In anticipation of impending extreme weather events, utilities assess their potential impact and pre-position resources. This proactive approach is crucial, as mobilizing resources can be time-consuming and may leave customers without power for extended periods. The speed and effectiveness (while putting safety of crews and customers as the top priority) with which human resources restore power after an outage is a key indicator of organizational resilience, and their performance should be systematically monitored and compared to previous events. However, there is currently no widely adopted metric for this purpose. The metrics describing emergency response need to be simple to promote adoption among utilities.

Grid automation technologies, such as feeder reconfiguration, customer restoration through alternative sources (like alternate feeders), sectionalization, energy storage systems, emergency resources, or microgrids, can also play a pivotal role in the restoration process~\cite{G1}-\cite{G2}. 
Additionally, a metric that tracks customer restoration progress during an event, and compares it to pre-established targets based on historical performance, can provide valuable insights into the progress and efficiency of restoration efforts. 
By monitoring the achievement of these targets across multiple events, utilities can assess their annual performance, identifying areas of improvement where restoration goals were not met.

\begin{figure}[thp]
\centering
\includegraphics[width=1.0\columnwidth]{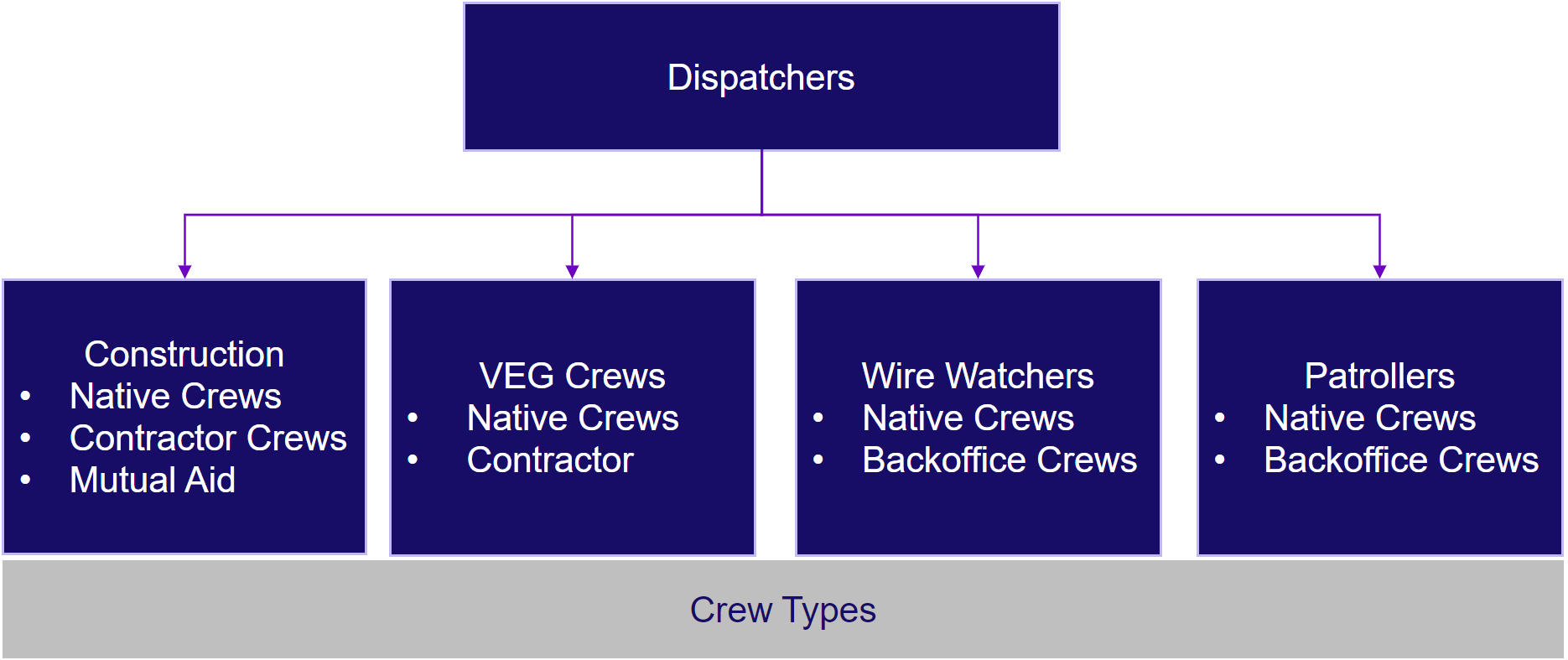}
\vspace{-1.5\baselineskip}
\caption{Crew Dispatching Hierarchy}
\vspace{-1.5em}
\label{crew_figure}
\end{figure}

It is crucial to understand the distribution utility restoration process \cite{Pandey1}. As shown in Fig.~\ref{crew_figure}, a team of dispatchers in the control center monitor the Outage Management System (OMS) and assign tickets to field crews to repair outages and restore service. This process relies on phone communication, which can sometimes result in incorrect crew dispatches. For example, if a tree brings down a power line, the sequence should be sending a wire watcher, then a tree crew, and finally a construction crew. Insufficient patrols or resources can lead to premature dispatch of construction crews, wasting time and labor without making progress on repairs.

We briefly indicate previous work calculating distribution system resilience metrics from observed data. 
Wei and Ji  \cite{WeiAM16} analyze distribution system resilience with data from Hurricane Ike with a non-homogeneous Poisson outage process arriving at a queue that repairs the outages to produce a dynamically varying restore process. 
Carrington  \cite{CarringtonPS21}  systematically extracts outage and restore processes and performance curves 
and their metrics from events in detailed utility outage data.
Kandaperumal \cite{KandaperumalSG22} develops metrics describing the threats, network topology, generation, loads, and restoration, and uses data from an Alaskan town to show how the metrics can be used to improve resilience planning and operation.  
Using EAGLE-I data scraped from web outage reports,
Ericson \cite{EricsonNREL22} extracts performance curves and frequency and duration statistics and
Abdelmalak \cite{AbdelmalakACCESS23} extracts the distributions of resilience metrics 
for extreme events that cross thresholds in the number of customers out.
Ahmad \cite{AhmadPS24} extracts baseline resilience metrics from 
detailed utility outage data, and introduces the rerunning history method to quantify the benefits of 
the overall effects of better restoration or hardening the grid for wind.

In this paper, we describe the challenges of emergency response and leverage and advance beyond \cite{CarringtonPS21,AhmadPS24} to define and extract new restoration metrics from utility data, including new metrics that describe crew performance, a topic that has been neglected in previous work.

\section{Emergency Response Effectiveness}

The efficiency of emergency response after extreme weather events depends on the event's severity and trajectory. Proactive measures, such as deploying crews to heavily impacted areas and optimizing restoration efforts, can improve the process. Utilities allocate crews based on forecasts and continuously update assessments during the event to allocate resources effectively. Despite best efforts, crew scheduling may not always be optimal due to the unpredictable nature of high-impact events. Crew managers prioritize outages with the highest customer counts and critical customers. Accurate forecasts are crucial; inaccuracies may require additional contractor crews and mutual assistance.

Restoration crews face travel challenges, road damage, and hazardous environments, making safety paramount. Repair tasks range from simple to complex, and supply chain issues can complicate the process. Also, event intensity does not always correlate with customer interruptions, especially in rural areas. Typically, crews are dispatched chronologically, but prioritizing high-customer outages and minimizing travel time is more effective. A metric to capture emergency response efficiency should be derived from available data, provide actionable insights, and be adaptable for widespread adoption.

\section{Extracting events \& metrics from utility data}
\begin{figure}[thpb]
\centering
\includegraphics[width=1.0\columnwidth]{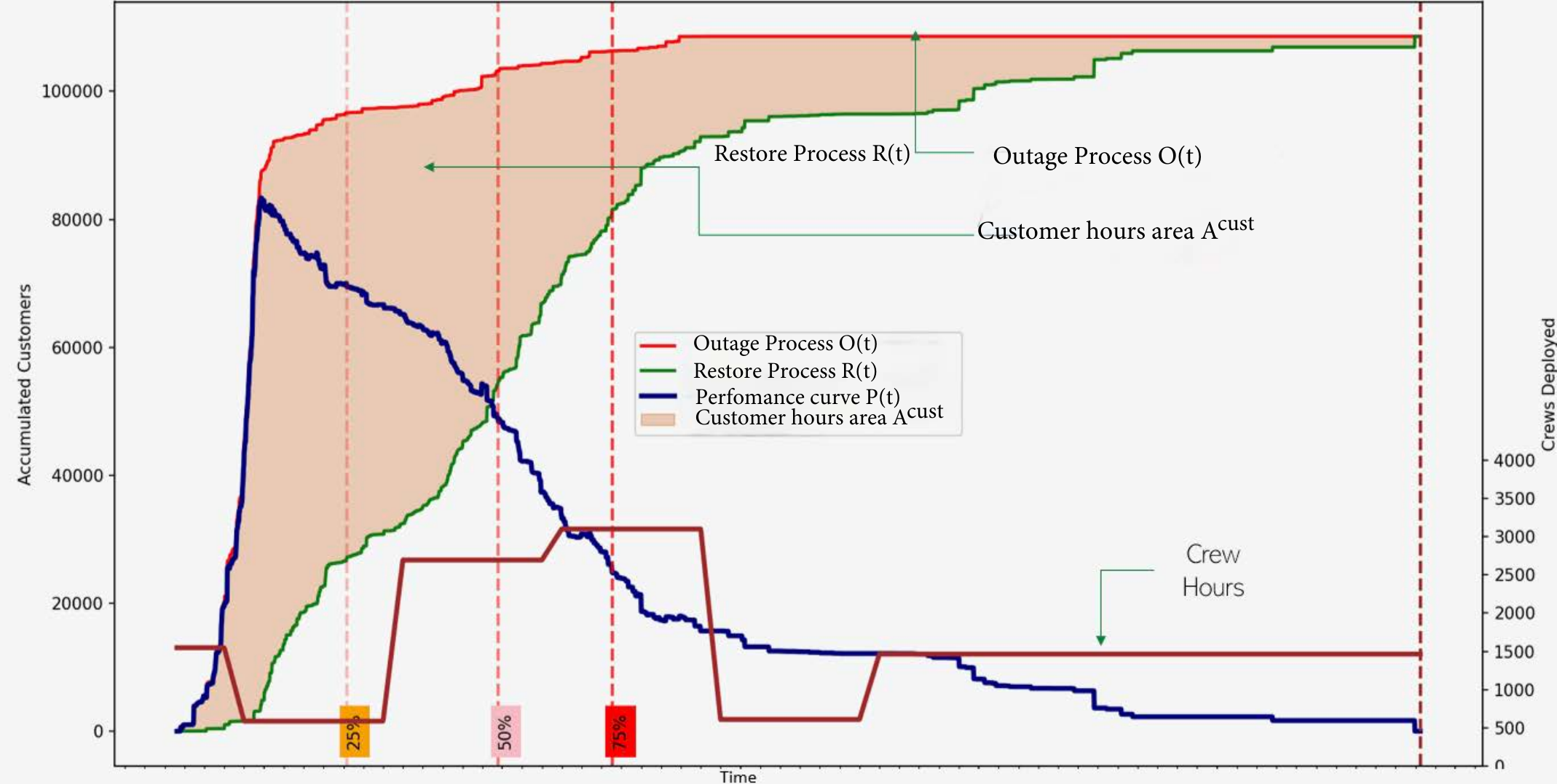}
\caption{Operational resilience processes}
\label{processesfigure}
\end{figure}

Utilities can readily process the outage data related to particular extreme weather events from their outage management system logs to capture the various processes that inform the operational resilience metrics. The accumulated customer outages form the outage process $O(t)$ and the accumulated restored customers form the restore process $R(t)$ as time $t$ increases. The performance curve $P(t)$ is the accumulated number of {\sl unrestored} customers. The area under the performance curve is equivalent to the area between the outage process and the restore process, which is quantified as the customer hours metric, ($A^{\text{cust}}$). Along with the crew deployment on the system by the hour during the event, Fig.~\ref{processesfigure} visualize the operational resilience processes.

Since operation resiliency analysis assesses system performance based on events, individual outages are grouped per extreme weather event. Since simultaneous outages occur when the grid is stressed, the grouping can be done by detecting the outages that overlap (that is, the outages that occur before all the previous outages are restored) to form the event \cite{CarringtonPS21}. The number of outages in an event ($n$) is a measure of the grid's strength relative to the storm intensity. The start time of an event $o_1$ is defined by an initial outage that occurs when all system components
are operational, and the end of the same event is defined by the first subsequent time $r_n$ when all the components
are restored. We write $o_1\le o_2\le\cdots\le o_n $ for the outage times in the order in which they occur and
$r_1\le r_2\le\cdots\le r_n $ for the restore times in the order in which they occur. The outages happen in the time interval
$[o_1,o_n]$ and the restores happen in the time interval $[r_1,r_n]$. In real outage data, the restores typically start before the outages end, so that these time intervals overlap.  



We write $c_k$ for the number of customers outaged at the $k$th outage.
The total number of customers outaged during the event is $n^{\rm cust}=c_1+c_2+\cdots+c_n$.  $n^{\rm cust}$ and $n$ are useful metrics describing the size of the event in terms of customer impact and utility impact respectively.
 
We form the curve of the customer outage process  $O^{\rm cust}(t)$, which is the accumulated number of customers outaged at time~$t$ during the event. And we form the curve of the customer restore process  $R^{\rm cust}(t)$, which is the accumulated number of customers restored at time $t$ during the event: 
\begin{align} O^{\rm cust}(t) = \sum_{o_k\le t} c_k ~\mbox{  and  }~
R^{\rm cust}(t) = \sum_{r_k\le t} c_k
\end{align}
As shown in Fig.~\ref{processesfigure},
the customer outage process $O^{\rm cust}(t)$ starts from zero at the beginning of the event and rises to the total number of customers out  $n^{\rm cust}$ during the event.  
The customer restore process $R^{\rm cust}(t)$ starts from zero shortly after the beginning of the event and rises to $n^{\rm cust}$ at the end of the event. The customer restore process, in effect, tries to catch up with the customer outage process, and the event ends when it does catch up and all the customers are restored.

\looseness=-1
The customer performance curve $P^{\rm cust}(t)$ tracks the accumulated number of {\sl unrestored} customers at time $t$ during the event as shown in Fig.~\ref{processesfigure}.
We have the relationship:
\begin{align}P^{\rm cust}(t) = O^{\rm cust}(t)-R^{\rm cust}(t).
\end{align}
\looseness=-1
This follows since the unrestored outages increase when customers are outaged and decrease when the customers are restored.  (Often $P^{\rm cust}(t)$ is plotted below the time axis and then it changes sign so that it is $R^{\rm cust}(t)-O^{\rm cust}(t)$ \cite{CarringtonPS21,DobsonPESL23,AhmadPS24}.)

 Instead of tracking the number of customers outaged or restored, we can similarly track the number of outages to obtain the outage process $O(t)$, which is the accumulated number of outages at time $t$ during the event, the restore process $R(t)$, which is the accumulated number of restores at time $t$, and the performance curve $P(t) = O(t)-R(t)$, which is the accumulated number of unrestored outages at time $t$.  

 Many useful metrics now follow from the dimensions and areas of the outage and restore processes. The metrics are summarized in Table \ref{tablemetrics}, including an example of their values for a particular storm.
 It is useful to divide the metrics into those describing the outage process and the strength of the grid relative to the storm, those describing the restore process and the recovery, and those describing the performance curve and the combined effect of the outage and restore processes.

{\bf Customer hours $A^{\rm cust}$:}
 Perhaps the most useful metric assessing the impact of the event on the customers is the total customer hours out $A^{\rm cust}$. $A^{\rm cust}$ combines the event size in terms of number of customers with duration. It turns out \cite{DobsonPESL23} that $A^{\rm cust}$  is also the area between the performance curve $P^{\rm cust}$ and the time axis and also the area between the outage process $O^{\rm cust}(t)$ and the restore process $R^{\rm cust}(t)$.
 We write $d_k$ for the duration of the $k$th outage. Then two ways to calculate $A^{\rm cust}$ are
 \vspace{-3mm}
\begin{align}
A^{\rm cust} &= \sum_{k=1}^nc_k d_k= \int_{o_1}^{r_n}|P^{\rm cust}(t)| dt 
\label{Acust}
\end{align}
Note that the  $A^{\rm cust}$ is the contribution of the outages in the event towards the numerator of \color{red}System Average Interruption Duration Index (SAIDI) \color{black} if the event is not excluded from SAIDI as a major event day. The resilience metric $A^{\rm cust}$ measures the customer hours lost in an event or class of events whereas SAIDI measures outages over the year.

	

\begin{table}[hbpt]
\normalsize
	\caption{\normalsize Resilience Metrics for an Event}
	\label{tablemetrics}
	\centering
	\setlength{\tabcolsep}{0.3em}
\begin{tabular}{rcr}
 \multicolumn{3}{c}{\scriptsize OUTAGE PROCESS METRICS} \\
 \hline
 number of outages &$n$ &{\small 850}\\
number of customers out&$n^{\rm cust}$  &{\small 88923}\\
outage duration (storm duration)&$o_n-o_1$  &{\small 29}\\
outage rate&$n/(o_n-o_1)$   &{\small 29.31}\\[1mm]
\multicolumn{3}{c}{\scriptsize RESTORE PROCESS METRICS}\\
 \hline
delay before start of restore&$r_1-o_1$ &{\small 0.567}\\
restore duration &$r_n-r_1$ &{\small 70}\\
duration to 95\% restore& $D^{\rm restore}_{95\%}$  &{\small 49}\\
customer restore rate&$n^{\rm cust}/(r_n-r_1)$  &{\small 1270.3}\\
outage restore rate&$n/(r_n-r_1)$   &{\small 12.14}\\
crew hours for restoration&$C^{\rm crew}$   &{\small 72264
}\\
Restoration Efficiency RE&$\log_{10}[C^{\rm crew}/n]$   &{\small 1.948}\\
[1mm]
\multicolumn{3}{c}{\scriptsize PERFORMANCE CURVE METRICS}\\
 \hline
event duration &$r_n-o_1$   &{\small 9}\\
max customers simultaneously out& $\max_t |P^{\rm cust}(t)|$    &{\small 32959}\\
max number simultaneous outages &$\max_t |P(t)|$    &{\small 501}\\
customer hours out and $P(t)$ area &$A^{\rm cust}$ &
{\small 753380
}\\
Area Index of Resilience AIR&$\log_{10}[A^{\rm cust}/n^{\rm cust}]$\hspace{-4mm} &{\small 0.928}\\
REPAIR\,=\,RE Plus AIR&RE\,+\,AIR   &{\small 2.876}\\
\end{tabular}
	\vspace{-3mm}
\end{table}
	\vspace{3mm}
{\bf Crew hours $C^{\rm crew}$:} One way to track the restoration effort of the utility is by the number of crews, measured by full-time equivalents (FTEs). The curve $C(t)$ is the number of crews deployed at time $t$. $C(t)$ can be recorded for each hour of the restoration. The total crew hours $C^{\rm crew}$ for the event can be evaluated by integrating $C(t)$ over the duration of the event, or, if $C(t)$ is recorded for each hour of the restoration, by the sum of the records.
Crew hours can include non-restoration activities like travel, preparation, and inventory restocking. Detailed task breakdowns during shifts may not always be feasible. Crew hours are influenced by storm intensity and duration, with limited staffing overnight further impacting hours. Minimizing crew hours for full restoration is key to better performance.



Given the basic metrics for each event described above, combinations of the basic metrics are defined in the following. 

{\bf Restoration Efficiency RE:}
The Restoration Efficiency is the logarithm of the average crew hours per outage restored:
\vspace{-0.9mm}
\begin{align}
\mbox{RE} = \log_{10}\bigg[\frac{C^{\rm crew}}{n}\bigg]
\end{align}
The logarithm is used because $C^{\rm crew}/n$ varies widely over events. 
More efficient crew deployments have lower RE.

{\bf Area Index of Resilience AIR}: The area index of resilience for an event is the logarithm of the event customer hours per customer, calculated as 
\vspace{-2mm}
\begin{align}
\mbox{AIR} = \log_{10}\bigg[\frac{A^{\rm cust}} {n^{\rm cust}}\bigg]
\end{align}
$A^{\rm cust}/n^{\rm cust}$ is a normalized form of the customer hours out that is less dependent on the event size and more dependent on the restoration performance. Indeed, for an average performance curve, $A^{\rm cust}/n^{\rm cust}$ is the average restore time minus the average outage time \cite{DobsonPESL23}.
The logarithm is used because $A^{\rm cust}/n^{\rm cust}$ varies over order of magnitudes for different events. 
A more efficient restoration has a lower AIR.

{\bf REPAIR = RE Plus AIR:}
REPAIR combines the customer hours for the average customer out and the average crew hours per outage restored by adding RE and AIR:
\begin{align}
{\rm REPAIR} = \mbox{RE + AIR} = \log_{10}\bigg[\frac{C^{\rm crew} }{n} \cdot \frac{A^{\rm cust}}{n^{\rm cust}}\bigg]
\end{align}
The crew effectiveness is captured in RE and a measure of the duration of the restoration process is captured in AIR.  A lower value of REPAIR indicates better overall performance. 
This first principles approach deriving REPAIR from more basic metrics should allow better quantification of the overall emergency response efficiency.

\color{red}
Our new metrics RE, AIR, and REPAIR are intended to quantify and balance restoration effectiveness and efforts. They do not directly describe all the individual aspects of system strength as other resilience metrics can usefully do, but $A^{\rm cust}$  quantifies a combined outcome of these individual strengths in real events.
While  $n$,  $C^{\rm crew}$,  $n^{\rm cust}$, and $A^{\rm cust}$ vary with the storm strength, the RE, AIR, and REPAIR metrics are normalized to reduce or eliminate this dependence. 
\color{black}


\section{Calculating  metrics from post-storm data}
\label{calculatemetrics}
As mentioned earlier, the crews need to be proactively placed near high impact areas and to be aware of repairs/restorations to be performed to bring the maximum number of customers back online in the minimum amount of time. 
The worked example below examines the different factors affecting human performance and computes the overall RE, AIR, and REPAIR. 
The example considers sample data from a set of storms going back to 2022 with some filtering performed to consider outages not impacted by the response efforts, removing crew hours dedicated to patrol and safety, etc. 
Metrics for the 9 storms are shown in Table \ref{stormmetricsvar}.

\begin{table}[hbpt]
\vspace{-2mm}
	\caption{Storm resiliency metrics}
	\label{stormmetricsvar}
	\centering
	\setlength{\tabcolsep}{0.3em}
    \vspace{-2mm}
\begin{tabular}{ crrrrrrr}
Storm	&$n$& $C^{\rm crew}$ & RE & $n^{\rm cust}$ &  $A^{\rm cust}$		&	AIR& REPAIR\\
\hline\\[-2.5mm]
1	&	1536	&	142172	&	1.966	&	176929	&	1135907	&	0.808	&	2.774	\\
2	&	1126	&	49549	&	1.643	&	107578	&	370417	&	0.537	&	2.180	\\
3	&	1267	&	42399	&	1.525	&	128132	&	282653	&	0.344	&	1.868	\\
4	&	216	&	31866	&	2.169	&	28724	&	31786	&	0.044	&	2.213	\\
5	&	2588	&	118405	&	1.660	&	208613	&	2221044	&	1.027	&	2.688	\\
6	&	850	&	75411	&	1.948	&	88923	&	753380	&	0.928	&	2.876	\\
7	&	457	&	30250	&	1.821	&	49497	&	91268	&	0.266	&	2.087	\\
8	&	347	&	30816	&	1.948	&	38053	&	80027	&	0.323	&	2.271	\\
9	&	1129	&	49443	&	1.641	&	111156	&	576270	&	0.715	&	2.356	\\
\vspace{-5mm}

\end{tabular}
\end{table}
\looseness=-1
We observe that the RE, AIR, and REPAIR scores provide some crucial insights. Using fewer resources to achieve faster restoration and containing the impact of customer interruptions generates better scores. Similarly, using disproportionate crew deployment for smaller storms results in lower scores, indicating that simply increasing the number of crews on the ground did not always improve the response in reducing the maximum number of customer interruptions, or the duration for full restoration, or both. 
Among the 9 storms analyzed in this study, storms 2, 3, and 9 had a similar impact on the system, with an average outage count of 1175 and average interruptions affecting 115,000 customers. These three storms provide a basis for a case study to compare how the metrics quantify the impact and thereby inform utilities about their emergency response efficiency.
\color{red}Comparing storms 2 and 9, we observe storm 2 outperforms storm 9 \color{black} in terms of the AIR score (0.537 vs. 0.715) and the REPAIR score (2.180 vs. 2.356). This difference is attributed to the crews’ ability to restore the system more quickly. The $A^{\rm cust}$ for storm 2 is significantly lower than that for storm 9, indicating a shorter time for full restoration.
Similarly, when comparing storms 2 and 3, the metrics show that storm 3 performed better than storm 2, despite comparable outages and customer interruptions. This can be explained by the fact that storm 2 required more crew hours to restore the system compared to storm 3 (49,549 vs. 42,399). Higher crew hours could indicate several factors: outages spread over a broader area, more labor-intensive and time-consuming repairs, difficult-to-access terrain, or possibly incorrect forecasts leading to suboptimal crew allocation and staging. 
\color{red}
The REPAIR metric captures this mismatch in outage impact versus crew efficiency between the two storms (1.868 vs. 2.180).
\color{black}




Table~\ref{stormmetricsvar} shows the variation of REPAIR with respect to the factors in the computation. 
Furthermore, additional factors should be considered to account for a) inability of restoration -- typically caused by safety concerns arising from extreme weather events where the crew cannot physically be on-site to restore the customers, and b) additional planned outages required to complete the restoration process.

\section{Estimating the impact of more crews}

\looseness=-1
Restoration speed can be improved by investing in additional repair crews, better inventory management, and optimized route scheduling. 
To estimate the benefits of faster restoration, we use the `rerunning history' approach \cite{AhmadPS24}. 
This historical rerun method provides insights into the improvements in resilience metrics that a proposed resilience investment would have produced if it had been implemented in the past. 
By relying on real data, this approach incorporates all the complex factors that have historically affected resilience. 
While it does not predict future outcomes, the historical rerun method is much simpler than forecasting with simulation models. 
Additionally, this approach can make a stronger business case for resilience investments to stakeholders, as it highlights benefits that would have directly impacted their past experiences, which can be more compelling than the hypothetical benefits of future simulations.

\looseness=-1
\color{red}
There are several ways that the REPAIR metric can help utilities assess the impact of crew investments on the expected improvement in resiliency. In this work, we illustrate three possible ways utilities can dispatch their crews to observe better REPAIR metrics and, thereby, improve emergency response.

\textbf{Increasing crew efficiency:}
One possible assumption is that an investment in restoration speeds up each restoration. 
Ideally, if the efficiency of the crew increases by 10\% and the number of crews is the same. Then each of the $n$ repairs would be completed 10\% faster.  
Since each repair is now completed 10\% faster (that is, each $d_k$, $k=1,2,\ldots ,n$ in (\ref{Acust}) would be 10\% smaller), event customer hours $A^{\rm cust}$ would also decrease by 10\%. This would lead to a reduction of approximately  $-\log_{10}0.9 =0.046$ in all the AIR and REPAIR metrics.

\textbf{Proactive dispatching of crews: }Adding crews early enough in the storm window would allow them to address outages that would otherwise be restored later, effectively pulling forward some restoration efforts that were delayed due to high outage volumes. As a result, the storm could be resolved a few hours earlier than in the base case, leading to zero crew hours beyond the restored time and thus achieving overall crew savings. The earlier restoration of outage tickets would also reduce the area under the performance curve, thus lowering the AIR metric.

\begin{figure}[t!]
\centering
\includegraphics[width=1.0\columnwidth]{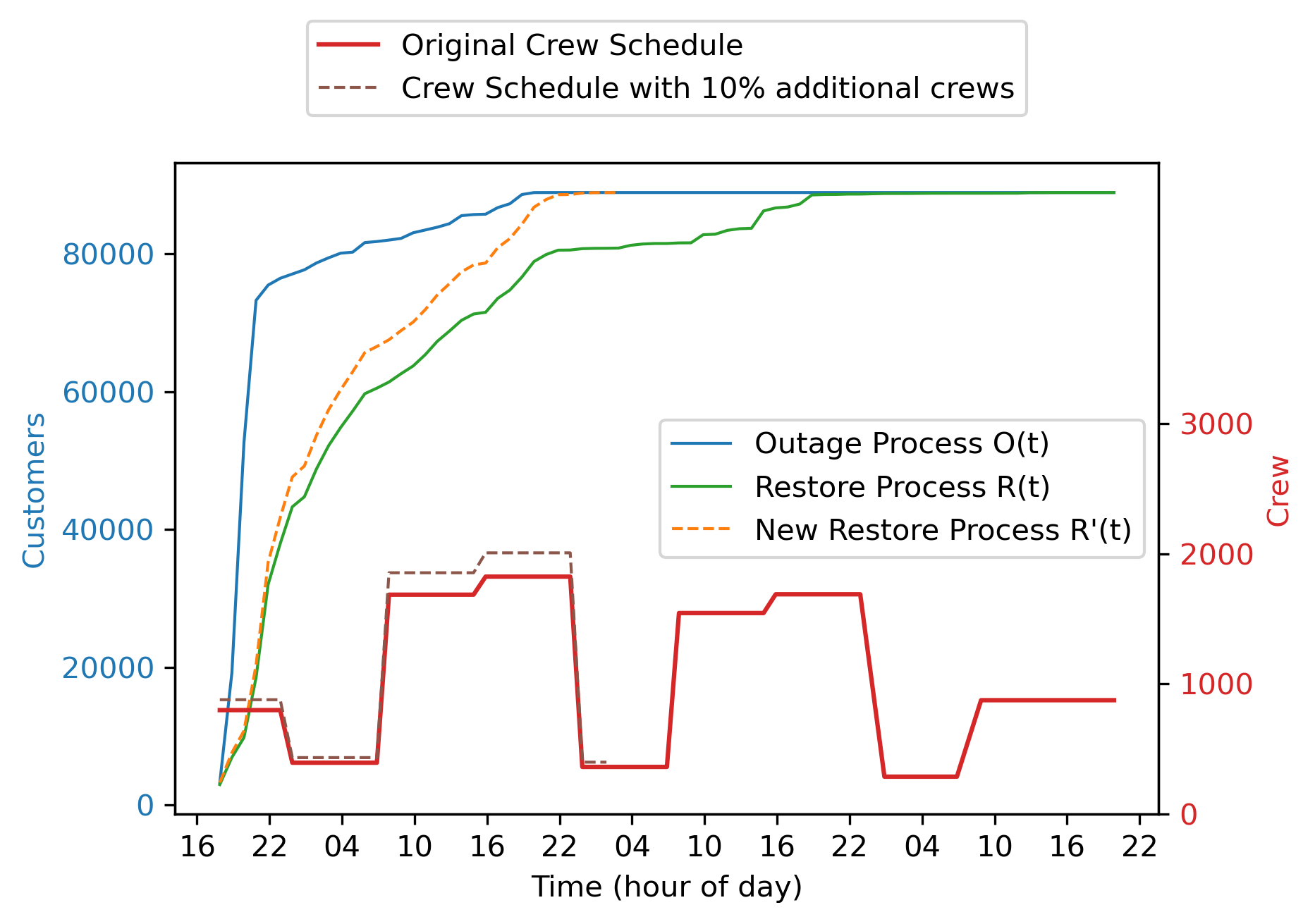}
\caption{Reduction in REPAIR with 10\% more crews}
\label{AIR_Sim}
\vspace{-5mm}
\end{figure}

\textbf{Increasing available crews: } Increasing the number of crews at the beginning of the storm can provide utilities the capacity to address more outages simultaneously and the flexibility in managing unexpected challenges during restoration efforts. To simulate this scenario and its impact on the REPAIR metrics, we work through re-running history for Storm 6 from Table II. In this approach, we added 10\% more crews at the beginning of the storm and assumed that a 10\% increase in crew hours at any given time would yield a corresponding 10\% decrease in restoration time per outage, resulting in a new restoration curve ($R'(t)$) as shown in Fig.~\ref{AIR_Sim}.

\looseness=-1
The simulation applies these assumptions directly to the outage logs used to process the accumulated customer interruptions $O(t)$ and the accumulated customer restorations $R(t)$. Thereby, for each outage $o_n$, the time taken to complete the corresponding restoration $r_n$ occurs 10\% faster. We denote the new restoration as $r'_n$. Simultaneously, the new restore process $R'(t)$ is computed to capture accumulated restored customers for each timestep $k$ (1 hour) using the formula:
\begin{align}
R'_{k}(t) = R'_{k-1}(t) + \bigg[\frac{r'_{k} - r_{k}}{r_{k}}\bigg]R'_{k-1}(t) 
\end{align}
Once the last outage is restored, all emergency crews are released from duty.Therefore, the simulation is terminated when $R'_{k}$ = $R_{k}$.
\color{black}
The simulation results are displayed in Table~\ref{stormmetrics1},
and we observe modest overall crew savings. However, depending on how additional crews are deployed, this could also lead to an increase in the RE metric. \color{red} We acknowledge that operational constraints such as the complexity of repairs, accessibility due to the nature of the event or geography of the location, availability of resources, and crew safety can impact this assumption and intend to pursue, in future work, a quantification of the corresponding decrease in restoration time per outage as crew numbers are increased. 
\color{black}

\begin{table}[t]
	\caption{Rerunning storm 6 resiliency metrics}
	\label{stormmetrics1}
	\centering
	\setlength{\tabcolsep}{0.3em}
    \vspace{-1mm}
\begin{tabular}{ crr }
Metric	&Base Case&Simulated Case \\
\hline
 RE&	1.948&	1.699\\
 AIR&0.928&	0.723\\
REPAIR&	2.876&	2.422
\end{tabular}
\vspace{-6mm}
\end{table}

\looseness=-1
These approximate estimates of the effects of more crews open the door to more elaborate estimates in future work. For example, as discussed in Section~\ref{calculatemetrics}, increasing the crews so much that repairs become saturated would decrease the crew efficiency. Different effects of starting restoration earlier (which may not be feasible until the storm ends) and starting at the same time but restoring faster can be readily calculated using the methods in \cite{AhmadPS24}. The overall effects of hardening can also be calculated \cite{AhmadPS24}.

\section{Conclusions}
The REPAIR metric combines storm performance with restoration resources, offering a comprehensive evaluation. This paper examines how increased crew staffing affects REPAIR outcomes. Future work should analyze the reduction in the performance curve area based on crew levels for optimal resource allocation. Considering individual utilities' operational constraints will enable pragmatic adoption. Tailoring the approach to specific needs and conditions will help utilities improve restoration efforts.

\bibliographystyle{IEEEtran}

\end{document}